\documentclass[12pt ]{article}
\usepackage{jheppub_kim}
 \usepackage{epstopdf}
\usepackage{amsmath}
\usepackage{amssymb}
\usepackage{dcolumn}
\usepackage{bm}
\usepackage{color}
\usepackage{epsfig}
\usepackage{amsfonts}
\usepackage{graphicx}
\usepackage{subfigure}
\usepackage{multirow}
\usepackage{xcolor}
 
\begin{document}
\title{Refined Thermodynamic Analysis of Perfect Fluid Dark Matter Black Holes in a Phantom Background} 
\author[a]{Jyotish Kumar,} 
  \author[b,c]{Sudhaker Upadhyay\footnote{Visiting Associate, Inter-University Centre for Astronomy and Astrophysics (IUCAA) Pune-411007, Maharashtra, India},}
  \author[d]{Himanshu Kumar Sudhanshu,}
  \author[e]{Yerlan Myrzakulov,}
  \author[e]{Kairat Myrzakulov,}
    \emailAdd{jyotishkumar7137@gmail.com}
 \emailAdd{sudhakerupadhyay@gmail.com; sudhaker@associates.iucaa.in}
 \emailAdd{himanshu4u84@gmail.com}
  \emailAdd{ymyrzakulov@gmail.com}
  \emailAdd{krmyrzakulov@gmail.com}
 \affiliation[a]{P.G. Department of Physics, Magadh University, Bodhgaya,  Bihar-824234, India }
   \affiliation[b]{Department of Physics, K.L.S. College, Magadh University, Nawada-805110,  India}
   \affiliation[c]{School of Physics, Damghan University, P.O. Box 3671641167,  Damghan, Iran}
   \affiliation[d]{Department of Physics, J. J. College, Magadh University, Gaya-823003,  India}
    \affiliation[e]{Department of General \& Theoretical Physics, L. N. Gumilyov Eurasian National University,  Astana, 010008, Kazakhstan}
\abstract{
In this study, we examine the thermodynamics of  black holes immersed in perfect fluid dark matter (PFDM) by employing the Misner-Sharp energy framework. We extend the analysis to include the thermal fluctuations of these black holes when size reduces to small size, redefining key thermodynamic variables such as pressure, volume, temperature, internal energy, and entropy within the context of PFDM. Using these newly defined variables, we systematically calculate the enthalpy, Gibbs free energy, and specific heat of PFDM black holes, incorporating perturbative thermal corrections. To assess the stability of these black holes, we perform a detailed graphical analysis of the specific heat as a function of the horizon radius, providing insights into the thermal stability of PFDM black holes. This comprehensive approach enhances our understanding of the thermodynamic properties and stability of black holes in the presence of dark matter.}
\maketitle

\section{Overview}
Astrophysical observations have provided strong evidence that supermassive black holes, surrounded by vast halos of dark matter, reside at the centers of large elliptical and spiral galaxies \cite{bg,bh}. When investigating black holes in realistic astrophysical contexts, it is essential to consider the impact of dark matter, which plays a significant role in the overall dynamics of these systems \cite{bi}. Moreover, recent experimental data suggests that the universe is experiencing an accelerated rate of expansion \cite{bj, bk}. To account for this observed acceleration, physicists have proposed new theoretical models, one of which introduces the concept of dark energy--an unknown form of energy characterized by negative pressure \cite{bl}. Numerous models have been put forth to explain the properties and behavior of dark energy \cite{bn}, with phantom energy being one of the most widely discussed and studied, due to its unique features and implications for both cosmology and black hole physics \cite{bn, bo, bp}.

In cosmological models, the assumption that the universe is homogeneous and isotropic at the background level leads to the adoption of the Friedmann-Lemaître-Robertson-Walker (FLRW) metric for characterizing the geometry of spacetime. Dark energy models, which aim to explain the accelerated expansion of the universe, can typically be categorized into two primary approaches: one that uses a fluid-based description \cite{bm} and another that is based on scalar field theory. In particular, phantom energy, a form of dark energy, is commonly modeled through a scalar field that exhibits negative kinetic energy in a homogeneous state. A notable and highly studied scenario in this context is the model of a spherically symmetric black hole embedded in a configuration of  PFDM  within a phantom dark energy framework. This model has garnered considerable analytical focus due to its intriguing implications for both black hole physics and cosmological phenomena \cite{bq}.

When black holes are treated as thermal systems, they do not fully adhere to the second law of thermodynamics unless the concept of entropy is integrated. It has been established that the maximum entropy of black holes is intrinsically linked to the area of their event horizon \cite{001,0II3}. This connection forms the basis for the development of the holographic principle, a key framework in modern theoretical physics \cite{003,004}. Extensive research has shown that the maximum entropy of black holes must be adjusted, leading to modifications of the holographic principle itself \cite{005,006}. These adjustments arise from the quantum nature of gravity and the thermal fluctuations around equilibrium. As the black hole shrinks due to Hawking radiation, these fluctuations become especially significant. The resulting corrections have been found to follow a logarithmic form at the first order, providing deeper insights into black hole thermodynamics \cite{0II5}. Recent investigations into these corrections have underscored their importance across a range of black hole types, including quasitopological black holes \cite{0015}, charged rotating black holes \cite{0016}, $f(R)$ black holes \cite{0018}, charged massive black holes \cite{0017}, black branes \cite{mir}, and Horava-Lifshitz black holes \cite{0019}. These studies have advanced our understanding of black hole physics, highlighting how quantum effects and thermal dynamics shape their behavior.

A detailed examination of the thermodynamic properties of PFDM black holes has revealed the necessity for corrections to these properties. In this study, we try to fill this gap and focus on deriving logarithmic corrections to the entropy of PFDM black holes, which arise due to thermal fluctuations. The first step involves computing these corrections and analyzing their impact by comparing the corrected and uncorrected entropy values as a function of the event horizon radius. Furthermore, by employing the first law of thermodynamics, we derive corrections to the internal energy of PFDM black holes and illustrate these modifications through comparative graphical analysis. Additionally, we compute the leading-order corrections to the free energy and investigate the effect of thermal fluctuations by comparing the corrected and uncorrected free energy as functions of the event horizon radius.

In addition to entropy and free energy, we analyze the corrected pressure of PFDM black holes and represent its variation with respect to the event horizon radius. We also compute the system's enthalpy, demonstrating its increase with the expansion of the horizon radius. Furthermore, we explore the influence of thermal fluctuations on the Gibbs free energy. Finally, we investigate the impact of these corrections on the specific heat of PFDM black holes, offering insights into the thermodynamic stability of this black hole thermodynamic system.

The paper is designed in the following ways. In section \ref{sec 2}, we briefly review and discuss the preliminary idea of a PFDM black hole solution. Thermodynamics of PFDM black hole and its correction due to thermal fluctuation are analyzed and discussed in section \ref{sec 3} and \ref{sec 4}. The stability of the PFDM black hole is addressed in section \ref{sec 5}. Finally, in section \ref{sec 6}, we discussed the conclusion and outcome of the paper.

In all our calculations, we adopt the metric signature \( (-, +, +, +) \) and use natural units with \( G = c = 1 \).

\section{PFDM Black Hole in Phantom Background} \label{sec 2}
 In a PFDM black hole framework, the presence of a non-homogeneous phantom field, minimally coupled to dark matter and gravity which interacts weakly with the phantom field. The action is formulated as \cite{br}
\begin{equation}
S = \int d^{4}x \sqrt{-g}\left[\frac{R}{2k^{2}}+\frac{1}{2}g^{\mu v}\partial_{\mu} \varphi \partial_{v}\varphi - V\left(\varphi \right)+ L_{m}+ L_{I}\right], \label{s}
\end{equation}
where \( k^2 = 8\pi G \),  \( V(\varphi) \) represents the potential of the phantom field. The term \( L_m \) corresponds to the Lagrangian describing the massive dark matter component within the galaxy, while \( L_I \) accounts for the interaction between dark matter and the phantom field. However, owing to the weak coupling between these two components, the interaction term \( L_I \) can be safely neglected for simplicity in the current analysis.

To investigate static and spherically symmetric solutions of the black hole system, the metric ansatz is given as  
\begin{equation}
 ds^{2} = -e^{v(r)}dt^{2} + e^{\mu\left( r\right) }dr^{2} + r^{2} d\theta^{2}+ r^{2}\sin^{2}\theta d\varphi^{2}. \label{a}
\end{equation}
Furthermore, the ansatz for the phantom field is specified as \( \varphi = \varphi(r) \). By varying the matter action in Eq. (\ref{s}) with respect to the metric tensor, we derive the stress-energy tensor \( T^\mu_{\ v} \) as shown below, where the dark matter Lagrangian \( L_m \) is modeled as a perfect fluid,
\begin{equation}
T^{t}_{t} = -\rho = \frac{1}{2}e^{-\mu}\varphi'^{2}- V(\varphi)- \rho_{_{DM}}, \label{b}
\end{equation}
\begin{equation}
T^{r}_{r} =\rho_{r} = \frac{1}{2}e^{-\mu}\varphi'^{2}- V(\varphi),\label{c}
\end{equation}
\begin{equation}
T^{\theta}_{\theta}= \rho_{\theta} = T^{\phi}_{\phi}= \frac{1}{2}e^{-\mu}\varphi'^{2}- V(\varphi),\label{d}
\end{equation}
The prime symbol \( ' \) indicates differentiation with respect to \( r \), while the dark matter stress-energy tensor is expressed as \( T_v^{\mu (DM)} = \text{diag}(-\rho_{_{DM}}, 0, 0, 0) \).

For the static situation, the Einstein field equations read 
\begin{equation}
g^{tt}R_{tt}- \frac{1}{2}R = e^{-\lambda}\left( \frac{1}{r^{2}}- \frac{\lambda'}{r}\right) -\frac{1}{r^{2}} = k^{2}T^{t}_{t},\label{e}
\end{equation}
\begin{equation}
g^{rr}R_{rr} -\frac{1}{2} = e^{-\lambda} \left( \frac{1}{r^{2}} + \frac{v'}{r}\right) -\frac{1}{r^{2}} = k^{2}T^{r}_{r},\label{f}
\end{equation}
\begin{equation}
g^{\theta\theta}R_{\theta\theta}- \frac{1}{R} =  \frac{e^{-\lambda}}{2}\left( v'' + \frac{v'2}{2}+ \frac{v'-\lambda'}{r}-\frac{v'\lambda'}{2}\right) =  k^{2}T^{\theta}_{\theta}=k^2 T_{\phi}^{\phi}.\label{g}
\end{equation}
In this analysis, the energy-momentum tensor represents the massive dark matter within the phantom field background. Our focus lies in obtaining solutions under the condition \( \lambda(r) = -v(r) \), which enforces the constraint \( T^{t}_{t} = T^{r}_{r} \) as derived from Eqs. (\ref{e}) and (\ref{f}). Under this constraint, the mass density of WIMPs is expressed as \( \rho_{_{DM}} = e^v \varphi'^2 \), as obtained from Eqs. (\ref{b}) and (\ref{c}). Notably, the kinetic energy term of the phantom field carries a negative sign, distinguishing it from standard scalar fields such as quintessence.

Assuming the stress-energy tensor satisfies the condition \( T^\theta_\theta = T^\phi_\phi = -\frac{1}{2} T^t_t \), we obtain the following result by combining Eq. (\ref{e}) and Eq. (\ref{g}):
\begin{equation}
e^v r^2 v'' + e^v r^2 v'^2 + 3e^vrv' + e^v-1 = 0.\label{h}
\end{equation}
By introducing the variable \( U = 1 - e^v \), Eq. (\ref{h}) can be recast in a more simplified form
\begin{equation}
r^2 U'' + 3rU' +U =0.\label{i}
\end{equation}
Solving this equation yields \( U = \frac{2M}{r} + \frac{\lambda}{r}\log\frac{r}{\lambda} \), leading to a non-trivial black hole solution given by \cite{br}
\begin{equation}
ds^2 = -\left( 1- \frac{2M}{r} -\frac{\lambda}{r}\log\frac{r}{\lambda}\right) dt^2 + \left( 1- \frac{2M}{r} -\frac{\lambda}{r}\log\frac{r}{\lambda}\right)^{-1} dr^2 +r^2\left( d\theta^{2}+ sin^{2}\theta d\phi^{2}\right).\label{j} 
\end{equation}
Here, \( M \) and \( \lambda > 0 \) are integration constants associated with the black hole's mass and the total energy density, representing the combined energy of dark matter and phantom dark energy, respectively. Moreover, the relationship between \( T_{\theta}^{\theta} \) and \( T_{t}^{t} \) imposes a constraint on the potential \( V(\varphi) \). 
\section{Thermodynamics of PFDM Black Holes: Insights and Analysis} \label{sec 3}
\begin{figure}  
\centering \includegraphics[width=0.5\textwidth]{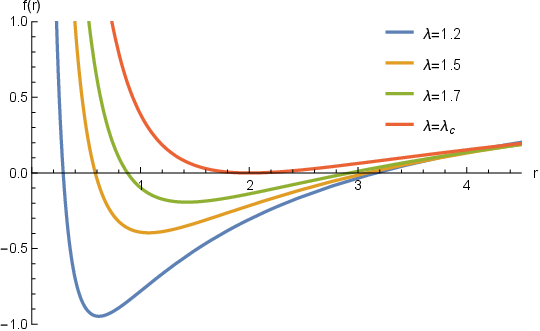}
\caption{Variation of metric function of PFDM black hole with $r$ for different values of $\lambda$.}
\label{metric}
\end{figure}

The prior section established that the black hole metric is described by equation (\ref{j}). The event (outer) horizon $r_{h} $ of the black hole can be calculated from the largest root of the metric function   $f(r)=0$ as 
\begin{equation}
f(r) =1- \frac{2M}{r} -\frac{\lambda}{r}\log\frac{\lambda}{r}= 1- \frac{r_{h}}{r} -\frac{\lambda}{r}\log\frac{r}{r_{h}},
\label{k}
\end{equation}
and same has been plotted in Fig. \ref{metric} with variation of $r$ for different values $\lambda$. 
It has been found that the black hole has both inner and outer horizon radius for depending on $\lambda$ and as the value of $\lambda$ increases, black hole approaches the extremal case for its critical value,  $\lambda_c = 2 M$.

From $f(r)$, the temperature of the PFDM black hole  can be easily calculated as \cite{br}
\begin{equation}
T_H= \frac{f'(r_{h})}{4\pi} = \frac{r_{h} -\lambda}{4 \pi r^{2}_{h}}.\label{l}
\end{equation}
As the above equation shows, since the temperature must be positive, we deduce that $ r_{h}\geq \lambda$ and the case $ \lambda= r_{h} $ corresponds to the external black hole.\\
The entropy is calculated as \cite{br}
\begin{equation}
S_{0} = \pi r_{h}^{2}.\label{m}
 \end{equation}
Since it is generally accepted that black holes significantly larger than the Planck scale possess entropy proportional to their horizon area, it becomes crucial to explore the primary corrections that arise as the size of the black hole decreases.
It has been found that the leading order corrections to the entropy of any thermodynamic system due to small
statistical fluctuations around equilibrium is logarithimic in nature \cite{0II5}. Considering the Planck scale size of PFDM black hole,  the first order correction  to entropy due to thermal fluctuation at the equilibrium is given by  \cite{0II5}
\begin{equation}
S =S_{0} - \frac{1}{2}\log S_{0}.
\label{n0}
\end{equation}
Now, we label second term by introducing a  label parameter $\beta$ also called as correction parameter which is equal to $1/2$ when fluctuation exists and vanishes for the system at
equilibrium \cite{hks}.
This gives 
\begin{equation}
S =S_{0} - \beta\log S_{0}.
\label{n}
\end{equation}
 By exploiting the relation in Eq.  (\ref{m})  and Eq.  (\ref{n}), the form of the corrected entropy due to thermal fluctuation for the PFDM black hole is obtained as
\begin{equation}
S= \pi r_{h}^{2} - \beta \log\pi r_{h}^{2}.\label{o}
\end{equation}
\begin{figure} 
\centering \includegraphics[width=0.5\textwidth]{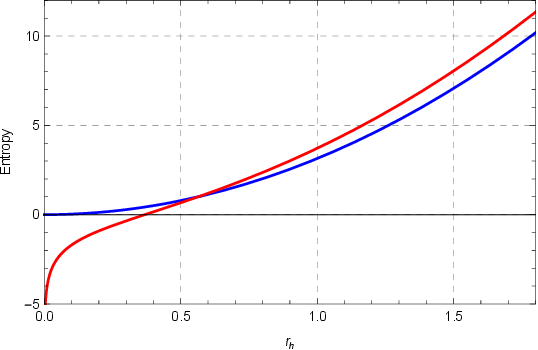}
\caption{ The   entropy versus horizon radius curve.  The blue curve represents the uncorrected entropy with \( \beta = 0 \), while the red curve shows the corrected entropy for \( \beta = 0.5 \).}\label{ai}
\end{figure}
As illustrated in Fig. \ref{ai}, the entropy of the black hole system remains positive in the absence of corrections. However, when thermal fluctuations are taken into account, a critical horizon radius emerges. Below this radius, the corrected entropy demonstrates a declining trend, exhibiting the unusual behavior of taking negative values.
\section{Refined Thermodynamic State Equations}\label{sec 4}
Here, we focus on computing the thermodynamic equations of the small black holes state by factoring in small thermal fluctuations within the system. With the entropy and temperature expressions, it becomes easy to compute other state equations. For example, the internal energy   $\left( E \right)$ of black hole can be found using the relation as
\begin{equation}
E = \int T_{H}dS.\label{p}
\end{equation}
By using equations (\ref{l}),  (\ref{o}), and (\ref{p}), we calculate the exact internal energy of the PFDM black hole as
\begin{equation}
E = \frac{1}{2}\left( r_{h} -\lambda \log r_{h}\right) + \frac{\beta}{2 \pi r_{h}}\left( 1+ \frac{\lambda}{2 r_{h}}\right).\label{q} 
\end{equation}
\begin{figure} 
\centering \includegraphics[width=0.5\textwidth]{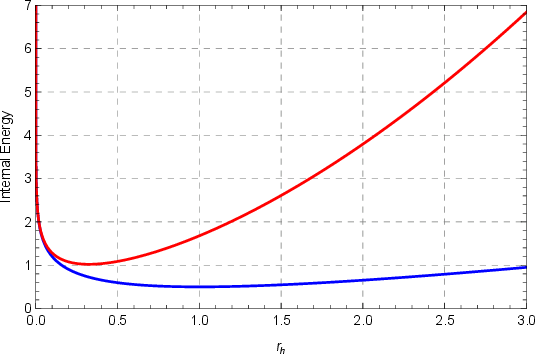}
\caption{Internal energy versus horizon radius plot of PFDM black hole. Here, blue line corresponds to $ \beta = 0 $ and red line corresponds to $ \beta =0.5 $, with $ \lambda = 1$.}
\label{aj}
\end{figure}
The variation of internal energy as a function of the horizon radius, as illustrated in Fig \ref{aj}. We observe that the system's internal energy is positive and increasing function with the horizon radius of the black hole. The correction term associated with thermal fluctuations directly contributes to the rise in internal energy of the black hole system.

\begin{figure} 
\centering \includegraphics[width=0.5\textwidth]{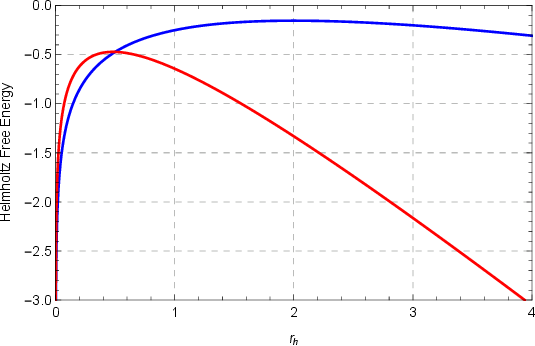}
\caption{ Helmholtz free energy versus horizon radius curve. The blue curve represents the case with \( \beta = 0 \) (uncorrected), while the red curve corresponds to \( \beta = 0.5 \) (corrected), with \( \lambda = 1 \).}
\label{ak}
\end{figure}

Now, the following formula can be used to compute the Helmholtz free energy of the PFDM black hole as
\begin{equation}
F = \int S d T_{H}.\label{r}
\end{equation}
By using equations (\ref{l}), (\ref{o}), and (\ref{r}), we calculate the free energy of the PFDM black hole as
\begin{equation}
F = \frac{\lambda}{2}  \log r_{h} -\frac{r_{h}}{4} + \frac{\beta }{4\pi r_{h}}\left[\left(\frac{\lambda - 2r_{h}}{r_{h}} \right) +\left(\lambda - 1\right)  \log \pi r ^{2}_{h}   \right].
\label{t}  
\end{equation}
The Helmholtz free energy behavior concerning the horizon radius is depicted in the plot Fig. \ref{ak}. 

Free energy of the black hole system is negative and found to be initially increases rapidly to a maximum value and then it becomes constant for the large horizon radius. Due to thermal fluctuations, corrected free energy is negative and rises quickly to a maximum value lower than the uncorrected one. After a particular value of horizon radius, it decreases more rapidly. For a specific horizon radius, it is independent of the correction parameter, $\beta$. 

\begin{figure} 
\centering \includegraphics[width=0.5\textwidth]{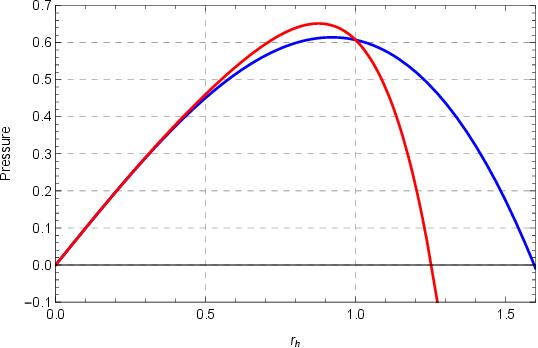}
\caption{Pressure versus horizon radius of the PFDM black hole. Here, blue line corresponds to $ \beta = 0 $ and red line corresponds to $ \beta = 0.5. $, with $ \lambda = 1$.}
\label{ah}
\end{figure}

From the area-entropy relation, we understand that the black hole entropy is proportional to the event horizon's area. Hence, the volume $\left(  V \right) $ of a PFDM black hole can be derived as
\begin{equation}
V= 4 \int S_{0} dr.\label{u}
\end{equation}
By using eqs. (\ref{m}) and (\ref{u}) volume of PFDM black hole found as 
\begin{equation}
V=\frac{4}{3}\pi r^{3}_{h}.\label{v}
\end{equation}
Considering the PFDM black hole as a thermodynamic system allows us to determine other thermodynamic parameters like pressure, $\left( P\right)$, as
\begin{equation}
P= - \dfrac{dF}{dV} = -\dfrac{dF}{dr_{h}}\dfrac{dr_{h}}{dV}.\label{w}
\end{equation}
Using the thermodynamic relations (\ref{t}) and (\ref{v}) into Eq.   (\ref{w}), we obtained the thermodynamic pressure of  the PFDM black hole as
\begin{equation}
P = \frac{\left( r_{h}-2\lambda\right) }{16 \pi r_{h}^{3}}- \beta\left[  \frac{\left( r_{h}-\lambda\right) }{8 \pi^{2}r_{h}^{5}}+ \frac{\left(\lambda -1 \right)\left( 2 - \log \pi r^{2}_{h}\right)  }{16 \pi^{2}r_{h}^{4}}\right].\label{x} 
\end{equation}
This relation (\ref{x}) describes the change in thermodynamic pressure experienced by the PFDM black hole due to thermal fluctuations. We explore the influence of thermal fluctuations on the thermodynamic pressure by the comparative plot of the equilibrium and the corrected pressures as functions of the event horizon radius of the PFDM black hole, which is illustrated in Fig. \ref{ah}.
We observe that the thermodynamic pressure of the PFDM black hole increases to a maximum value and then decreases sharply after that. The maximum value of thermodynamic pressure of this black hole system increases due to thermal fluctuation. For a particular horizon radius, it is independent of the correction parameter, $\beta$. 
Enthalpy ($H$) plays an essential role in the thermodynamics system of the black hole, which represents the total heat content within the black hole system and is mathematically defined as
\begin{equation}
H= E + P V.\label{y}
\end{equation}
By using the expression of corrected internal energy in Eq.   (\ref{q}) and pressure (\ref{x}) along with volume in Eq.   (\ref{v}),  the corrected enthalpy of the PFDM black hole is obtained from relation (\ref{y}) as
\begin{equation}
H = \frac{7r_{h}}{12} -\frac{\lambda\left( 1+3 \log r_{h}\right) }{6} + \beta\left[ \frac{\left( 4 r_{h} +5\lambda\right) }{12 \pi r^{2}_{h}} - \frac{\left(\lambda -1 \right)\left(2 - \log \pi r^{2}_{h} \right)  }{12 \pi r_{h}}\right].
\label{z} 
\end{equation}

\begin{figure} 
\centering \includegraphics[width=0.5\textwidth]{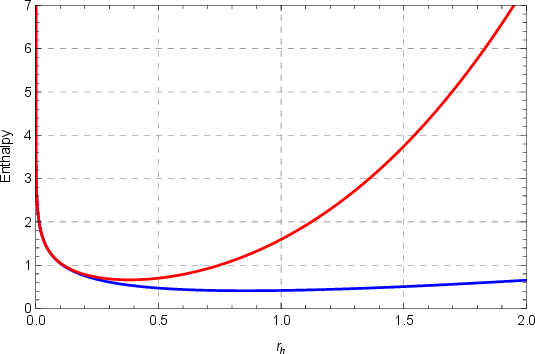}
\caption{Enthalpy versus horizon radius of PFDM black hole. Here, blue line corresponds to $\beta = 0 $ and red line corresponds to $ \beta = 0.5$, with $ \lambda = 1$.}\label{ag}
\end{figure}

To investigate the impact of thermal fluctuations on the enthalpy, we plot its variation with the horizon radius of the PFDM black hole as displayed in Fig. \ref{ag}. 
We observe that the enthalpy of the PFDM black hole increases more rapidly due to the correction parameter, $\beta$, for the large value of horizon radius. 

The Gibbs free energy represents the maximum amount of work that can be extracted from a thermodynamically isolated system, achievable only under a fully reversible process. To account for thermal fluctuations and determine the corrected Gibbs free energy (\( G \)) for the PFDM black hole, we employ the following definition:
\begin{equation}
G = F + P V.\label{aa}
\end{equation}
By substituting the corrected free energy from Eq. (\ref{t}), the pressure from Eq. (\ref{x}), and the volume from Eq. (\ref{v}) into Eq. (\ref{aa}), we derive the expression for the corrected Gibbs free energy of the PFDM black hole as follows
\begin{equation}
G = \frac{1}{6}\left(3 \lambda \log r_{h} -r_{h} -\lambda \right) +\frac{\beta}{12 \pi r^{2}_{h}}\left[\left( 3\lambda -8 r_{h} +2\right) +\left(3\lambda r_{h} +\lambda -3r_{h}-1\right) \log \pi r^{2}_{h} \right].\label{ab} 
\end{equation}
\begin{figure}
\centering \includegraphics[width=0.5\textwidth]{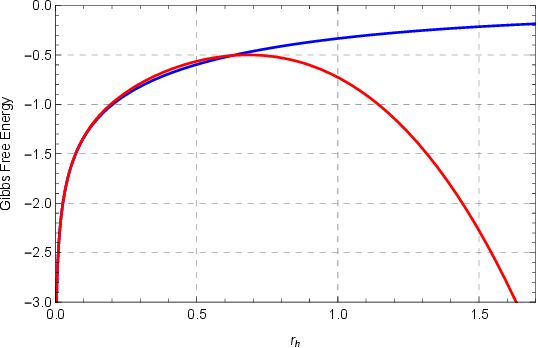}
\caption{Gibbs free energy versus horizon radius of the PFDM black hole. Here, blue line corresponds to $ \beta =0 $ and red line corresponds to $ \beta = 0.5. $, with $ \lambda = 1$.}\label{af}
\end{figure}
To investigate the impact of thermal fluctuations behaviour on the Gibbs free energy, we plot its variation with the horizon radius as shown in Fig. \ref{af}, where it is observe that the Gibbs free energy has consistently negative value. The correction parameter does not significantly affect Gibbs free energy in the small black hole region. Still, for the larger horizon radius, the correction parameter, $\beta$, has decreased the value of Gibbs free energy significantly. After a particular horizon radius, where Gibbs free energy is found to be indecent of the correction parameter, $\beta$, it decreases sharply. 
\section{Stability of PFDM Black Hole}\label{sec 5}
\begin{figure}
\centering \includegraphics[width=0.5\textwidth]{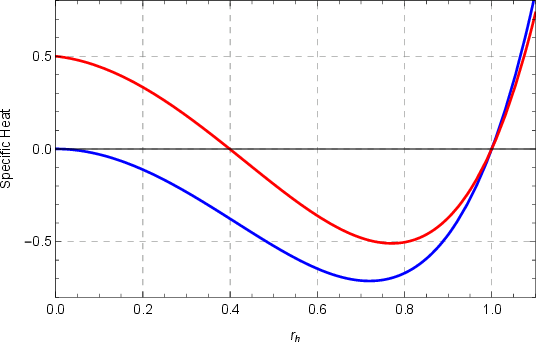}
\caption{Specific heat versus horizon radius of the PDFM black hole. Here, blue line corresponds $ \beta = 0 $ and red line corresponds to $ \beta =0.5.  $, with $ \lambda = 1$.}\label{ae}
\end{figure}

To evaluate the stability of the PFDM black hole, we compute its specific heat, including the effects of small thermal fluctuations on it. The stability or instability of the black hole is dictated by the sign of the specific heat, a positive sign of specific heat signify the stability of the black hole. By using standard relation, the corrected specific heat $(C_{v}) $ is calculated as
\begin{equation}
C_{v} = \dfrac{dE}{dT} = \dfrac{dE}{dr_{h}} \dfrac{dr_{h}}{dT}.\label{ac}
\end{equation}
By using Eq.  (\ref{q}) and Eq.  (\ref{l}), the corrected specific heat of the PFDM black hole found as 
\begin{equation}
C_{v} = 2 \pi r_{h}^{2}\left( \frac{r_{h} -\lambda}{2\lambda - r_{h}}\right)- 2\beta \left(\frac{r_{h}- \lambda}{2\lambda - r_{h}} \right).
\label{ad}  
\end{equation}
To study the effect of thermal fluctuation on the specific heat of the PFDM black hole, we plot the variation of specific heat with the horizon radius as depicted in Fig.
 \ref{ae}. 
 
 In the absence of thermal fluctuation, the PFDM black holes are unstable for the small region of the horizon and stable in the large horizon region. Due to the thermal fluctuation, we observe that the PFDM black hole changes state of its stability twice as the horizon radius increases, first from stable to unstable state and then to again stable one similar to its previous case, i.e. as in the absence of correction parameter. Hence, the correction parameter, $\beta$ has affected the stability of the PDFM black hole in small horizon region.

\section{Concluding Remarks}\label{sec 6}
We have explored how thermal fluctuations affect the PFDM black hole. Our results show that thermal fluctuations around equilibrium introduce a logarithmic adjustment to the entropy of the black hole. Additionally, we studied how thermal fluctuations influence the behavior of the thermodynamic parameters of the PFDM black hole. From this, we have derived the correction term for the entropy of the PFDM black hole. We observed that while the effect of the correction parameter does not play significance role in the entropy of the large black holes, it has a substantial impact on the entropy for smaller black holes regions. Furthermore, the corrected internal energy for the PFDM black hole was estimated using the first law of thermodynamics, and we also calculated the corrected Helmholtz free energy, which has been affected by thermal fluctuations in the large black hole regions. 

In our study, we have also calculated the thermodynamic pressure and volume of the black holes. The graphical analysis shows that pressure rises as the horizon radius increases to a specific horizon radius, after which it starts to decline with a further increase in the horizon radius. For the correction term, we observe that the pressure increases to its peak value and then rapidly decreases with an increase in the horizon radius. Furthermore, we derived the corrected expression for the enthalpy. We found that the enthalpy increases slowly with the horizon radius, but the effect of thermal fluctuations leads to an exponential rise as the horizon radius expands.   

To assess the stability of the black hole, we studied the corrected form of the specific heat for the PFDM black hole due to the thermal fluctuation. We found that the equilibrium-specific heat initially decreases and then increases with the expansion of the event horizon radius. Nonetheless, due to thermal fluctuations, the specific heat exhibits two changes, from positive to negative, which indicates a shift from stable to unstable, and then from negative to positive, indicating a return from unstable to stable region as the horizon of the PFDM black hole increases. Without thermal fluctuation, the black hole takes only one transition from unstable to stable region with increased horizon radius. Exploring the effect of non-perturbative corrections on the PFDM black hole thermodynamics will be a promising topic for future research.
\section*{Acknowledgement}
This research was funded by the Science Committee of the Ministry of Science and Higher Education of the
Republic of Kazakhstan (Grant No. AP23487178).


\begin{thebibliography}{99}
   \bibitem{bg} K. Akiyama et al.[Event Horizon Telescope], Astrophys. J. Lett. 875, L1 (2019). 
   \bibitem{bh} V.C. Rubin, W. K. Ford, Jr. and N. Thonnard, Astrophys. J. 238, 471 (1980).
    \bibitem{bi} V. V. Kiselev. class. Quant. Grav. 20, 1187 (2003).
 \bibitem{bj} S. Perlmutter  et al.[Supernova Cosmology Project], Astrophys. J. 517, 565  (1999).
 \bibitem{bk} A. G. Riess et al. [Supernova Search Team], Astron. J.
 116, 1009 (1998).
 \bibitem{bl} P. J. E. Peebles and B. Ratra, Rev. Mod. Phys. 75, 559  (2003).
  
 \bibitem{bn} K. Bamba, S. Copozziello, S. Nojiri and S. D. odintsov Astrophys. Space Sci. 342, 155 (2012). 
\bibitem{bo} R. R. Caldwell, Phys. Lett. B 545, 23 (2002).
\bibitem{bp} R. R. Caldwell, M. Kamionkowski and N. N. Weinberg, Phys. Rev. Lett. 91, 071301 (2003).
\bibitem{bm} S. Nojiri, S. D. Odintsov and S. Tsujikawa, Phys. Rev. D 71 (2005) 063004.
\bibitem{bq} S. Haroom, M. Jamil, K.Jusufi, K. Lin and R. B. Mann, Phys. Rev. D 99,   044015 (2019).


\bibitem{001} S. W. Hawking, Nature 248, 30 (1974).
  \bibitem{0II3} S. W. Hawking,   Commun. Math. Phys. 43, 199 (1975).
\bibitem{003} L. Susskind, J. Math. Phys. 36, 6377 (1995).
\bibitem{004} R. Bousso, Rev. Mod. Phys. 74, 825 (2002).
\bibitem{005} D. Bak and S. J. Rey, Class. Quant. Grav. 17, L1 (2000).
\bibitem{006} S. K. Rama, Phys. Lett. B 457, 268 (1999).
 \bibitem{0II5}S. Das, P. Majumdar and R. K. Bhaduri,   Class. Quant. Grav. 19, 2355 (2002). 
\bibitem{0015}  S. Upadhyay, Phys. Lett. B 775, 130 (2017).
\bibitem{0016} S. Upadhyay, Gen. Rel. Grav. 50, 128 (2018).
  
\bibitem{0018} S. Upadhyay, S. H. Hendi, S. Panahiyan and B. E. Panah, Prog. Theor. Exp. Phys.
2018, 093E01 (2018).
\bibitem{0017} S. Upadhyay, S. Soroushfar and R. Saffari, Mod. Phys. Lett. A 36, 2150212 (2021).
 
\bibitem{mir} B. Pourhassan and M. Faizal, JHEP 10 (2021) 050.
\bibitem{0019} B. Pourhassan, S. Upadhyay, H. Saadat and H. Farahani, Nucl. Phys. B 928, 415
(2018).


\bibitem{br} X. Liang, Y.P. Hu, C. H. Wu and Y. S. An, arXiv:2308.00308v2 [gr-qc].


\bibitem{hks} H.K. Sudhanshu, S. Upadhyay, D.V. Singh,  et al., Int J Theor Phys 61, 248 (2022). 

\end{thebibliography}
\end{document}